\documentclass[aps,prl,preprint,unsortedaddress,showkeys,nofootinbib]{revtex4-1}

\usepackage{amssymb,amsmath,amsfonts}

\usepackage{graphics}
\usepackage{graphicx}
\usepackage{epstopdf}
\usepackage[pdftex]{hyperref}
\usepackage{longtable}

\usepackage{bm}

\usepackage{color}
\usepackage{setspace}
\usepackage{multirow}

\usepackage[ruled,vlined,linesnumbered]{algorithm2e}

\usepackage{color, colortbl}
\definecolor{Yellow}{rgb}{1,1,0}
\definecolor{Grey}{rgb}{.87,.87,.87}
\definecolor{Purple}{rgb}{.8,.0,1.0}
\definecolor{Crimson}{rgb}{.86,.08,.23}

\begin{document}

\title{Inferring hierarchical structure of spatial and generic complex networks through a modeling framework}

\author{Stanislav Sobolevsky
\footnote{To whom correspondence should be
addressed: sobolevsky@nyu.edu}}
\affiliation{New York University}



\date{\today}

\begin{abstract}
\begin{it}
Our recent paper [Grauwin et al. Sci. Rep. 7 (2017)] demonstrates that community and hierarchical structure of the networks of human interactions largely determines the least and should be taken into account while modeling them. In the present proof-of-concept pre-print the opposite question is considered: could the hierarchical structure itself be inferred to be best aligned with the network model? The inference mechanism is provided for both - spatial networks as well as complex networks in general - through a model based on hierarchical and (if defined) geographical distances. The mechanism allows to discover hierarchical and community structure at any desired resolution in complex networks and in particular - the space-independent structure of the spatial networks. The approach is illustrated on the example of the interstate people migration network in USA.
 \end{it}
\end{abstract}

\keywords{Complex networks | Community detection | Network science}

\maketitle

\section*{Introduction}
Community detection is commonly used to discover the underlying structure of complex networks \cite{fortunato2016community} including networks of human mobility and interactions. Communities of the least in many cases were found to be spatially cohesive and largely consistent with the regional structure at municipal \cite{kang2013exploring}, country \cite{Ratti2010GB, blondel2010regions, Sobolevsky2013delineating, amini2014impact} and global \cite{hawelka2014geo, belyi2017global} scales \cite{Ratti2010GB, blondel2010regions} enabling applications to regional delineation. 

A recent paper of the author and collaborators \cite{grauwin2017identifying} demonstrated that the network community and hierarchical structure is also important for modeling human mobility and interactions - the study introduced a concept of the network hierarchical distance and showed that by itself this discrete parameter might have higher predictive power compared to a detailed knowledge of the spatial layout of the interacting individuals. 

However the way hierarchical distance is defined in \cite{grauwin2017identifying} relies on the knowledge on the nested network community structure. There exist a large body of community detection approaches, including straightforward heuristics \cite{Hastie2001ElementsOfStatisticalLearning,GN} and optimization techniques based on the maximization of various objective functions including modularity as the most commonly used one \cite{newman2004,newman2006}, as well as Infomap description code length \cite{Rosvall01052007InformationTheoretic, Infomap}, block model likelihood \cite{Newman2011Stochastic,Newman2011Efficient,Bickel2009Nonparametric, Decelle2011BlockModel, Decelle2011BlockModelAsymptotics, Yan2012ModelSelection}, and surprise \cite{Aldecoa2011Deciphering}. Specific modularity maximization algorithms include \cite{NewmanPRE2004, CNM2004VeryLargeNetworks, newman2004, newman2006, Sun2009, leuven, simulatedAnnealing,Good2010PerformanceOfModularity, Duch2005CElegans, LeeCSA, combo}. Some algorithms allow to fit the entire hierarchical structure of the network at once \cite{peixoto2014hierarchical}. But clearly all those algorithms require the detailed knowledge on the network itself to find the community structure and the hierarchical distance.

Sometimes an existing structure underlying the set of interacting objects, like a regional structure of the country, could be used instead and the associated hierarchical distance also turns out to be useful for modeling purposes \cite{grauwin2017identifying}. This rises a following question - what is the best hierarchical distance one could use for modeling complex networks? Which one is the most consistent with the network and the proposed modeling framework? Clearly, the most streightforward way to define such a distance would be to infer it as the part of the model inference framework. This will not only supplement the model but will also reveal the network hierarchical structure implied by the inferred hierarchical distance and community structure in particular. 

Moreover an appropriate model for the spatial networks (i.e. having their nodes located in geographical space with implied proximity relation in addition to the network links) will allow to distinguish the impact of the network hierarchy from the impact of geographical distance. A first attempt of finding space-independent communities in spatial network has been undertaken in \cite{expert2011uncovering} through a modification of the modularity objective function. The present paper will provide an approach for defining space-independent hierarchical structure for the spatial networks able to explain as much network information as possible in addition to what could be explained by the geographical distance alone.

This paper provides a general definition of the hierarchical distance for the complex network and constructs a network modeling framework inspired by the standard gravity model \cite{krings2009urban} and its further modifications like local normalization \cite{grauwin2017identifying} incorporating both - hierarchical and geographical (if provided in case of a spatial network; could be omitted otherwise) distances. The framework is further provided to infer the most suitable hierarchical distance for the network having the highest utility for the model. In particular it enables community detection in both - spatial and generic - complex networks at any given resolution.

\section{The hierarchical structure and hierarchical distance in complex networks}
The network community structure can be described with a binary relation $p(a,b)=0$ if $a$ and $b$ belong to the same community and $p(a,b)=1$ otherwise. This function $p$ is symmetrical (meaning that $p(b,a)=p(a,b)$) and the relation $p(a,b)=0$ is also transitive, i.e. if  $p(a,b)=0$ and $p(b,c)=0$ then $p(a,c)=0$. The least could be equivalently expressed through a sort of a triangle inequation $p(a,b)\leq \max(p(a,c),p(b,c))$. Also $p(a,a)=0$ for any $a$.

Instead of this binary relation introduce a more general non-negative real-valued function $h(a,b)$ such that $h(a,a)=0$ for any $a$, $h(b,a)=h(a,b)\neq 0$ for any $a\neq b$, and for any $a,b,c$ the following triangle inequation holds:
\begin{equation}
h(a,b)\leq \max(h(a,c),h(b,c)).
\label{trian}
\end{equation}

{\bf Definition 1.} Call the non-negative real valued function $h$ of pairs of network nodes satisfying the above conditions a {\it valid hierarchical distance} over the network.

A valid hierarchical distance is supposed to reflect the hierarchical level of relation characterizing the hierarchical structure of the network. Specifically it defines how close the two nodes are in the implied network hierarchy. At any hierarchical level $t$, the corresponding community partition $p_t(a,b)$ can is implied by assigning $p_t(a,b):=0$ if $h(a,b)\leq t$ and $p_t(a,b)=1$ otherwise. The inequation (\ref{trian}) ensures the transitivity of this relation.

From (\ref{trian}) it also follows that if $h(a,c)\neq h(b,c)$ then $h(a,b)=max(h(a,c),h(b,c))$ (otherwise if say $h(a,c)>h(b,c)$ and $h(a,b)<max(h(a,c),h(b,c))=h(a,c)$ then $h(a,c) \leq \max(h(a,b),h(b,c))<h(a,c)$).
In short there can be no triangle $a,b,c$ with $h(a,b), h(a,c), h(b,c)$ all different.

\section{Hierarchical model for generic complex networks}
Ideal hierarchical structure should perfectly describe the network connectivity, perhaps with respect to the heterogeneity of node strength. E.g. this might be represented as a model that for certain incoming and outgoing strength $w^{in}, w^{out}$ associated with each node and mapping $f$ will represent the network edge weights $e(a,b)$ as 
\begin{equation}
e(a,b)=w^{out}(a)w^{in}(b)f(h(a,b)).
\label{genmodel}
\end{equation}

Of course for real networks (\ref{genmodel}) might not hold precisely as the edge weights might not be perfectly consistent with the definition of the valid hierarchical distance, specifically with (\ref{trian}). Then problem of inferring the most suitable valid hierarchical distance could be expressed as fitting non-negative weights $w^{out}$, $w^{in}$, a valid hierarchical distance $h$ and a decreasing function $f$ (clearly, the further away in the network hierarchy the nodes are the weaker should be their connection) in order to minimize the error of the model (\ref{genmodel}). The least could be expressed as
\begin{equation}
EH_1=\sum_{a,b, a\neq b}\left(A(a,b)-m(a,b)\right)^2\to\min
\label{eh1}
\end{equation}
or
\begin{equation}
EHL_1=\sum_{a,b}\left(A(a,b)-m(a,b)\right)^2\to\min
\label{ehl1}
\end{equation}
depending on whether we care about fitting loop edges or not, where the modeled value $m(a,b)=w^{out}(a)w^{in}(b)f(h(a,b))$ according to (\ref{genmodel}). The above distance corresponds to the statistical inference of the model
$$
e(a,b)\sim N(m(a,b),\sigma^2)
$$
for constant $\sigma$.

Alternatively thinking of the network edges weights as a cumulative amount of interaction happening between the two nodes (e.g. this is the case for the aggregated networks of human mobility or interaction between locations) a Poisson model  
$$
e(a,b)\sim P(m(a,b)),
$$
where the model (\ref{genmodel}) is meant to represent the expected intensity of interactions between pairs of nodes over a period of time and the network edges are the actually observed amounts of interaction. For large $e(a,b)$ the Poisson model could be approximated by a normal model with equal mean and variance
$$
e(a,b)\sim N(m(a,b),m(a,b)).
$$

Then the max-likelihood criteria is equivalent to 
\begin{equation}
EH_2=\sum_{a,b, a\neq b}\frac{(e(a,b)-m(a,b))^2}{m(a,b)}\to \min
\label{eh2}
\end{equation}
or
\begin{equation}
EHL_2=\sum_{a,b}\frac{(e(a,b)-m(a,b))^2}{m(a,b)}\to \min
\label{ehl2}
\end{equation}
depending on whether we care about fitting loop edges or not. 

In the following inferences the loop edges will be omitted since they are not related with the hierarchical structure of the network and building the model capable of fitting them as well is beyond the primary objective of the present study. This way the criteria (\ref{eh1}) or (\ref{eh2}) will be used. Also without loss of generality one could consider a specific decreasing function $f(h)$ instead of fitting it, while the values of $h$ are left to be fit (as the definition of the valid hierarchical distance does not depend on the specific values of $h$ but only on their order relations for different edges). In the following inferences assume $f(h):=1/h-1$, while $h\in [0,1]$ (as we omit the loop edges the infinite value of $f$ for them do not affect the inferences).

\section{Hierarchical model for spatial networks}
What makes the spatial networks a particular case of the complex networks is the presence of geographical distance $d(a,b)$ between the arbitrary nodes $a,b$, while this distance is often related to the network connections \cite{krings2009urban}. And while hierarchical distance might often have even higher predictive power than the geographic distance might have \cite{grauwin2017identifying}, an adequate model should probably use both. This will also allow to distinguish the particular impact of the hierarchical distance which goes beyond what geographical distance could explain.

The model (\ref{genmodel}) could be generalized as following to account for the effect of the geographical distance:
\begin{equation}
e(a,b)=w^{out}(a)w^{in}(b)f(h(a,b))g(d(a,b)).
\label{spatmodel}
\end{equation}

In case of a generic complex network where the geographical distance is not provided it is sufficient to let $d(a,b):=1$ and $g(1):=1$ which will reduce (\ref{spatmodel}) to (\ref{genmodel}).

The criteria (\ref{eh1})-(\ref{ehl2}) proposed for fitting (\ref{genmodel}) are equally applicable for fitting the above model (\ref{spatmodel}). The only difference is that the function $g$, characterizing the impact of geographical distance is also subject to fit. One can use some specific form for $g$ like a power law $g(d)\sim d^{-\gamma}$ or $g(d)\sim exp(-\gamma d)$ or perform discrete binning the distances and fit the value of $g(d)$ separately for each bin $d$. Also the function $g$ could be first fit without $f(h)$ (let $h(a,b):=0.5$ for all $a\neq b$) and then already known $g$ will be used to fit $h$, or, alternatively, the functions $g$ and $h$ could be fit together (as suggested before let $f:=1/h-1$). The implementation for the last, most general, approach is proposed in the next section.

\section{Inference strategy}

In order to infer the weights $w^{out}$, $w^{in}$, $g(d)$ and the hierarchical distance $h$ in the model (\ref{spatmodel}) or (\ref{genmodel}) (assigning $g:=1$ in this case) given the set of network edge weights $e(a,b)$ the following framework could be proposed.\\
0. Select an initial random valid network hierarchy h.\\
1. Let available set S be all the nodes.\\
2. For each available hierarchy level $H$ from lower to higher ones: \\
2a. For all pairs of nodes $a,b$ consider increasing or decreasing $h(a,b)$ together with all the related edges needed to be adjusted in order to satisfy (\ref{trian}); fit the optimal new value of $h(a,b)$.\\
2b. Find the best adjustment from 2a providing the largest improvement to the selected objective function (\ref{eh1}) or (\ref{eh2}) and implement it.\\
2c. Adjust $w^{out}$, $w^{in}$, $g(d)$ using the formulas below.\\
3. If any improvement has been during the last series of steps 2 - repeat 2. Otherwise - stop.

Each of the $w^{in}$, $w^{out}$, $g(D)$ could be fit using the framework below (assuming the rest are fixed):
$$
0=\frac{\partial L}{\partial w^{out}(a)}\sim \sum_b \frac{\partial m(a,b)}{\partial w^{out}(a)}\left[2\frac{(A(a,b)-m(a,b))}{m(a,b)}+\left(\frac{(A(a,b)-m(a,b))}{m(a,b)}\right)^2\right]
$$$$
0=\sum_b \frac{\partial m(a,b)}{\partial w^{out}(a)}\left[\left(\frac{A(a,b)}{m(a,b)}\right)^2-1\right].
$$$$
w^{out}(a)=\sqrt{\sum_b \left(\frac{A(a,b)^2}{w^{in}(b)g(d(a,b))f(h(a,b))}\right)/\sum_b w^{in}(b)g(d(a,b))f(h(a,b))}
$$
Same way
$$
w^{in}(b)=\sqrt{\sum_a \left(\frac{A(a,b)^2}{w^{out}(a)g(d(a,b))f(h(a,b))}\right)/\sum_b w^{out}(a)g(d(a,b))f(h(a,b))}
$$
$$
g(D)=\sqrt{\sum_{a,b, d(a,b)=D} \left(\frac{A(a,b)^2}{w^{out}(a)w^{in}(b)f(h(a,b))}\right)/\sum_{a,b, d(a,b)=D} w^{out}(a)w^{in}(b)f(h(a,b))}
$$

The formulas above could be used in the iterative manner to adjust each of the quantities to fit during the step 2c, for example: adjust all $w^{in}$ at once first, then all $w^{out}$ at once, then each $f(D)$ one by one, then balance the overall magnitude of $w^{in}$ and $w^{out}$ by rescaling them, then repeat the procedure until no further improvement is possible. This algorithm converged pretty quickly. 

Finally, adjusting the $h(a,b)$ in 2a requires analyzing all the related triangles and defining a set of edges to be adjusted together ensuring that (\ref{trian}) is not violated. The adjustment could be done in a similar manner as the above formulae for $g(D)$, ensuring the new value stays on an interval $[H,min_{h>H}h]$ or $[\max_{h<H}(h),H]$.


\section{Application to the migration network of USA}

\begin{figure}[h]
\centering
\includegraphics[width=0.6\textwidth]{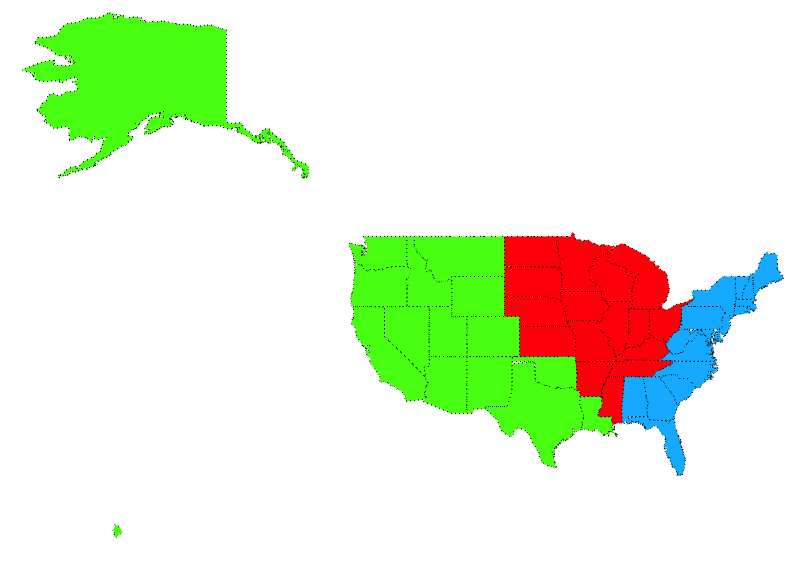}
\caption{\label{fig::USmod}Migration network partitioning using modularity optimization.}
\end{figure}

Now illustrate the approach on the example of US state-to-state migration network \cite{Census} where states are the nodes and edges between them are weighted by the volume of migrants from one state residing in the other. The optimal partitioning obtained through modularity maximization using the Combo algorithm \cite{combo} splits the country into just three major areas - East Coast, West Coast and the rest of the country as shown on the Figure \ref{fig::USmod}. 

\begin{figure}[h]
\centering
\includegraphics[width=0.6\textwidth]{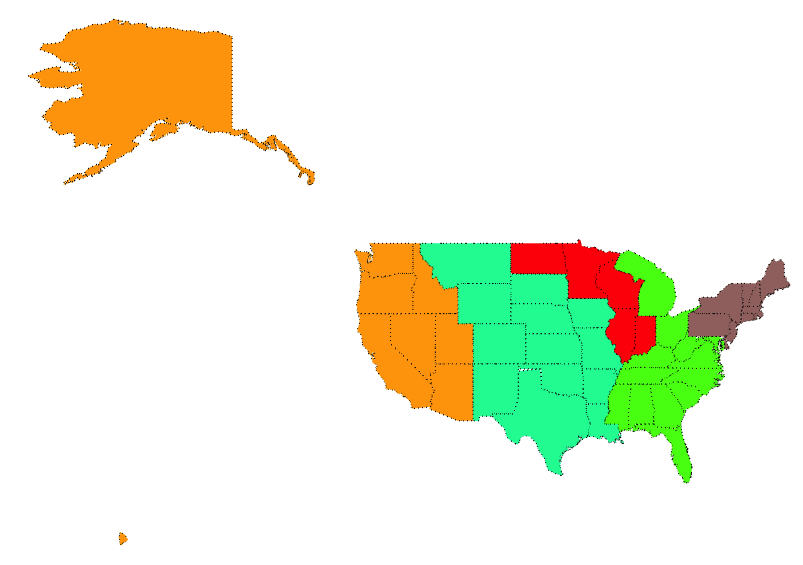}
\caption{\label{fig::USpart5}Section of the hierarchy for migration network, implying partitioning into 5 communities}
\end{figure}

\begin{figure}[h]
\centering
\includegraphics[width=0.6\textwidth]{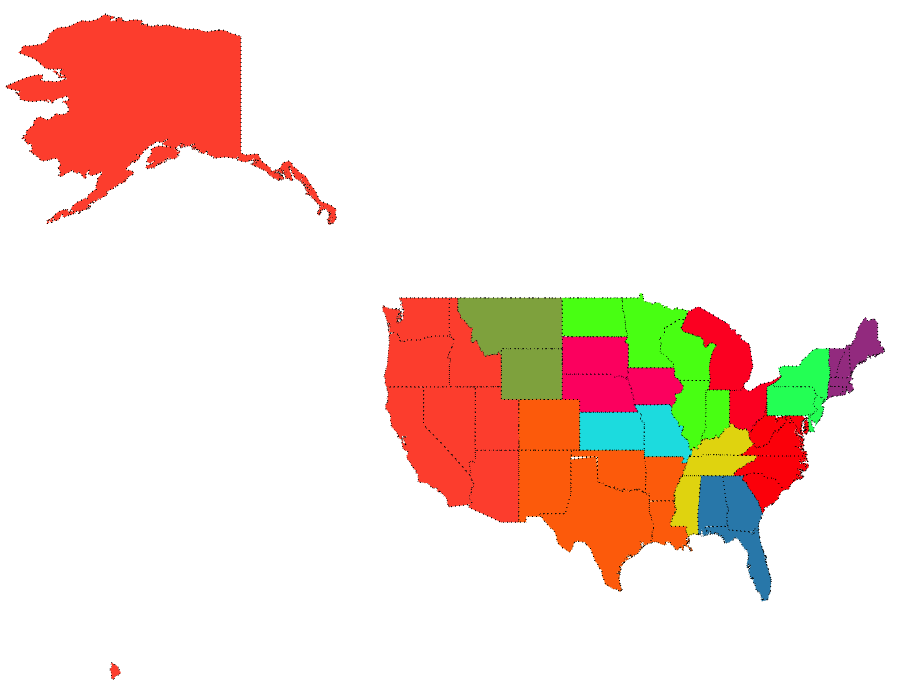}
\caption{\label{fig::USpart12}Section of the hierarchy for migration network, implying partitioning into 12 communities}
\end{figure}

A hierarchy for the network most suitable in terms of the model (\ref{genmodel}) with optimization criteria (\ref{eh2}) allows to produce partitioning at any desired resolution as a section of the hierarchy at a certain hierarchical level. E.g. partitioning in 5 and 12 communities are shown on the figures \ref{fig::USpart5}, \ref{fig::USpart12} providing a regional structure of the country capturing more fine-grained spatial patterns. Similarly to the modularity partitioning those produce spatially cohesive connected regions.

\begin{figure}[h]
\centering
\includegraphics[width=0.6\textwidth]{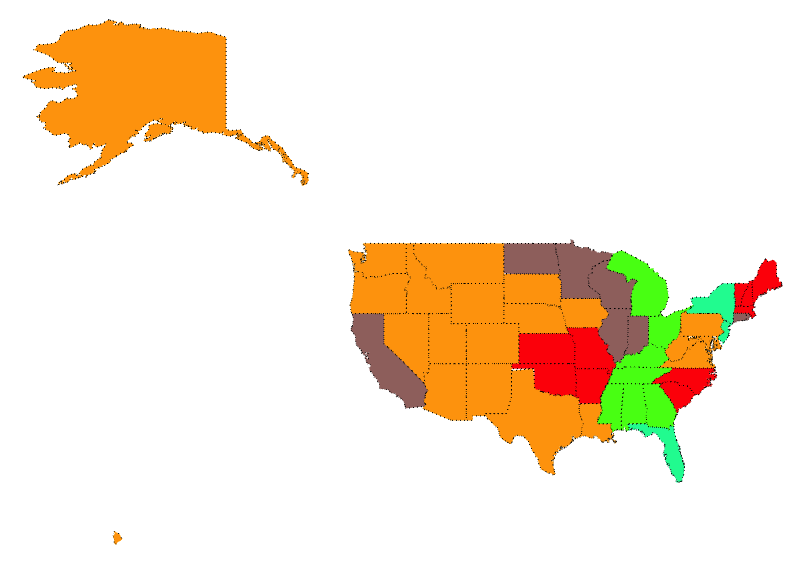}
\caption{\label{fig::USpartD}Section of the distance-independent hierarchy for migration network, implying partitioning into 5 communities}
\end{figure}

On the other hand a space-independent hierarchy most suitable in terms of the model (\ref{spatmodel}) with optimization criteria (\ref{eh2}) is capable of capturing important links between distant states as shown on the figure \ref{fig::USpartD}.

\section{Conclusions}
The present proof-of-concept pre-print provided a general definition of the hierarchical distance for the complex network. For both spatial or generic complex networks a network model has been proposed incorporating both - hierarchical and, in case of spatial networks, also geographical distances. A framework is provided to infer the most suitable hierarchical distance for the network having the highest utility for the proposed model. In particular, the proposed framework enables community detection in both - spatial and generic - complex networks of any given resolution. An approach is illustrated on the example of the interstate people migration network in USA, constructing regular and space-independent hierarchy and community structure for this network.

\bibliographystyle{apsrev}

\section{Figures}

\end{document}